\documentclass[aps, amsmath, twocolumn, showpacs, preprintnumbers,superscriptaddress,sort&compress, floatfix, amssymb]{revtex4-1}

\usepackage{graphicx}
\usepackage{rotating}
\usepackage{dcolumn}
\usepackage{bm}
\usepackage{color}
\usepackage{mathptmx, textcomp}
\usepackage[latin1]{inputenc}
\usepackage{braket}

\begin{document}

\author{Vladimir S. Melezhik}
\affiliation{Bogoliubov Laboratory of Theoretical Physics, Joint Institute for Nuclear Research,
Dubna, 141980 Dubna, Russia}
\author{Peter Schmelcher}
\affiliation{Zentrum f\"ur Optische Quantentechnologien, Universit\"at Hamburg, Luruper Chaussee 149, 22761 Hamburg, Germany}

\title{Multichannel Effects near Confinement-Induced Resonances in Harmonic Waveguides}

\date{\today}


\begin{abstract}
We analyze the impact of multichannel scattering in harmonic waveguides on the positions and widths of confinement-induced
resonances for both isotropic and anisotropic transversal confinement. Multichannel scattering amplitudes and transmission coefficients
are calculated and used to characterize the resonant behavior of atomic collisions with varying anisotropy. A mechanism is established
which leads to a splitting of the confinement-induced resonance in the presence of anisotropy.
\end{abstract}

\maketitle

\section{Introduction}

Confinement-induced resonances (CIRs), originally predicted  in the seminal work of Olshanii \cite{Olshanii1998} and more recently
observed experimentally for both bosons \cite{Kinoshita2004,Paredes2004,Haller2009} and fermions \cite{Gunter05},
have attracted a great deal of attention during the past few years. The immediate reason is that they represent a
valuable tool for the control of the atomic interactions thereby allowing to enter and probe the regime of strongly
correlated bosonic or fermionic many-body systems. Beyond this, changes of the transversal confinement potential possess
direct consequences for the scattering behavior of the atoms in the waveguide such that a multitude of binary resonance profiles
are accessible. Examples for the variety of scattering properties in waveguides are multi-channel confinement-induced 
resonances \cite{Saeidian2008}, resonant molecule-formation processes and center of mass coupling effects \cite{Melezhik2009,Peano2005}
as well as a dual CIR \cite{Melezhik2006} which leads to complete quantum suppression of scattering.

A recent experiment \cite{Haller2010} on CIRs in transversally anisotropic waveguides has shown the necessity of an adequate theoretical approach 
for describing collisional processes in confined
geometries of one- and two-dimensional character. A main result of this experimental work \cite{Haller2010} was the observation of a splitting
of the CIR-related loss signals in the presence of a transversal anisotropy. Two recent attempts to explain this splitting have not been
successful \cite{Peng2010,Zhang2}. The corresponding studies represent investigations of the pseudo-potential scattering with a single
open transverse channel \cite{Olshanii1998} in the case of tightly confining waveguides.
Both the width of the CIR and the full multichannel character of the problem were not taken into account.
On the other hand it has been shown that the non-separability of the center of mass and internal motion
for atomic scattering in anharmonic (transversally isotropic) waveguide potentials changes the resonance picture 
qualitatively and new resonances (anharmonicity-induced resonances (AIRs))  are occur where molecular excited center of mass states cross the
threshold \cite{Peano2005,Kestner2010}. Very recently the coupling of the center of mass excitations
in anharmonic isotropic and anisotropic confining potentials to the ground state has been analyzed in great detail in ab initio calculations \cite{Saenz2011}.
There a very good agreement of the AIR splitting with the distance between the maxima of the atomic loss in the experiment \cite{Haller2010} was found. 
However, it has not been analyzed what happens with the ``harmonic'' part of the investigated two-body spectrum in the trap and whether the ``harmonic'' CIR agrees with 
previous results \cite{Olshanii1998,Peng2010,Zhang2,Bergeman2003,LowDSystems}. 

In previous works  \cite{Olshanii1998,Bergeman2003,LowDSystems,Peng2010,Zhang2} the specific ratio $a_\perp/a_s=1.4603...$
(where $a_\perp = \sqrt{\hbar/(\mu\omega_\perp)}$,$a_s$,$\mu$ and $\omega_\perp$ are the harmonic oscillator length, scattering length in free space,
reduced atomic mass and harmonic oscillator frequency, respectively) yielding the position of the CIR was defined
by the zero $Im\{f_0(a_\perp/a_s)\}=0$ of the scattering amplitude $f_0$ in the transversal ground state in the zero-energy limit.
This point also corresponds to the absolute minimum of the transmission
coefficient $T_0(a_\perp/a_s)=|1+ f_0(a_\perp/a_s)|^2\rightarrow 0$ which represents an important scattering observable
near the CIR \cite{Olshanii1998,Bergeman2003,LowDSystems,Peng2010,Zhang2}.
In the present work, we explore a situation where not only the ground transversal channel but also excited channels contribute to the
scattering process in the waveguide and demonstrate the importance of multichannel scattering effects in the resonant region.

Within our description of few-channel ultracold scattering in confined geometries
we observe a mechanism which leads to the splitting of the CIR under the action of a transversal anisotropy of the waveguide.
It is based on the fact that the total transmission coefficient $T$, which we define as a sum of partial coefficients $T_i$ emerging from the
different transverse (ground and excited) states labeled by $i$,
\begin{equation}
T=\sum_i W_i T_i
\end{equation}
averaged over the initial populations $W_i$, possesses its main contribution near the CIR from the first excited
transversal state and not from the ground state.  This follows from the observation that $T_0$ possesses a much deeper as well as
broader transmission well around its minimum compared to the wells (minima) of $T_i(i\neq 0)$ for the excited states near the CIR \cite{Saeidian2008}.
Thus, even for low-energy pair collisions the scattering properties near the CIR location are not determined by the
partial coefficient $T_0$, where $T_0\rightarrow
0$, but by the behavior of the coefficients $T_i\neq 0$ in excited states $i\neq 0$.
Employing this mechanism we find a splitting of the minimum of the total
transmission coefficient (1) in an anisotropic trap. The splitting is a consequence of the different dependencies of the partial
coefficients $T_i(a_\perp/a_s)$ on $a_\perp/a_s$ for different states in the resonant region. A necessary prerequisite of the
appearance of this splitting effect is therefore the occupation of excited transversal states in the waveguide.
We emphasize, that the present investigation is performed for harmonic traps, i.e. we investigate the influence of the 
anisotropy of a harmonic transversal confinement on the CIRs \cite{Olshanii1998,Bergeman2003,LowDSystems,Peng2010,Zhang2} opposite 
to the above-mentioned works that explore the effects of anharmonicity \cite{Peano2005,Kestner2010}
and anisotropy \cite{Saenz2011}.

In the following section we describe the computational approach to our multichannel scattering problem in the confined quasi-1D geometry. 
The corresponding results are discussed in the third section. The short fourth section is devoted to a discussion of the mechanism leading
in the harmonic waveguide with transverse anisotropy to the splitting of the CIRs. Finally we provide a brief conclusion.

\section{Multichannel scattering problem in anisotropic harmonic
waveguides}

To calculate the partial transmission coefficients (the single index $i$ is here replaced by the double index $n_1,n_2$ indicating
the quantum numbers belonging to the different transversal degrees of freedom of the waveguide)
\begin{equation}
T_{n_1,n_2} =
\sum_{n_1',n_2'}\frac{k_{n_1',n_2'}}{k_{n_1,n_2}}\mid\delta_{n_1,n_1'}\delta_{n_2,n_2'}+f_{n_1,n_2}^{n_1',n_2'}\mid^2\,\,\,,
\end{equation}
describing the transmission probability from the initial transverse state $(n_1,n_2)$ to
all possible final states $(n_1',n_2')$ in the course of the collision of identical bosons in a harmonic waveguide with the transverse
trapping potential $\frac{1}{2}\mu(\omega_1^2x^2 + \omega_2^2y^2)$, we solve the multichannel scattering problem for the 3D Hamiltonian
\begin{equation}
H(x,y,z) =-\frac{\hbar^2}{2\mu}\triangle_{\bf r} +\frac{1}{2}\mu\omega_1^2x^2
+\frac{1}{2}\mu\omega_2^2y^2 + V(r) \,\,\,,
\end{equation}
depending on the relative variables ${\bf r}=(x,y,z)$, with the asymptotic scattering wave function
for $\mid z\mid\rightarrow +\infty$
$$
\psi_{n_1,n_2}({\bf r})=\cos(k_{n_1,n_2}z)\phi_{n_1,n_2}(x,y) + \sum_{n_1',n_2'=0}^{m_1,m_2}
f_{n_1,n_2}^{n_1',n_2'}
$$
\begin{equation}
\times \exp\{ik_{n_1',n_2'}\mid z\mid\}\phi_{n_1',n_2'}(x,y)\,\,\,.
\end{equation}
Here the matrix elements $f_{n_1,n_2}^{n_1',n_2'}(E)$  of the scattering amplitude
describe the transition from the initial channel with the
transverse energy
$E_\perp^{(n_1,n_2)}=\hbar [\omega_1(n_1+\frac{1}{2})+\omega_2(n_2+\frac{1}{2})]$ and the
relative longitudinal momentum
$\hbar k_{n_1,n_2} =\sqrt{2\mu(E-E_\perp^{(n_1,n_2)})}=\sqrt{2\mu E_\parallel }$ ($E_\parallel$ being the relative longitudinal collision energy
and $E$ the total energy) to the final open channel $(n_1',n_2')$  with $E=E_\perp^{(n_1',n_2')}+E_\parallel '$. $\phi_{n_1,n_2}(x,y)$ are the
eigenfunctions of the 2D harmonic oscillator corresponding to the
eigenvalues $E_\perp^{(n_1,n_2)}$. The latter are degenerate
with respect to the quantum number $n=n_1+n_2$ in an isotropic trap $\omega_1=\omega_2=\omega_\perp$ and $E_\perp^{(n_1,n_2)}\rightarrow E_\perp^{(n)}=\hbar\omega_\perp(n+1)$.
The asymptotic wave function (4) is explicitly symmetric with respect to the exchange of the atoms.

In order to solve the multichannel scattering problem (3,4) we extend the approach developed in \cite{Melezhik1991,Melezhik2003} for scattering in three dimensions:
the expansion over the spherical harmonics on a grid (two-dimensional discrete-variable representation) is replaced by the expansion over the product states
$\phi_{n_1,n_2}(x,y)=\varphi_{n_1}(x)\varphi_{n_2}(y)$ where $\varphi_{n_i}$ are the eigenfunctions of the 1D harmonic oscillator.
The calculations are performed with the finite-range Gaussian approximation $V(r)=-V_0 \exp\{-r^2/r_0^2\}$ for the interparticle interaction
with $V_0$ chosen such that we have one weakly-bound state in the potential well.
We do expect that all effects shown below are to a large extent independent of the chosen interaction
potential $V(r)$. As an independent check, we have verified that the results obtained for the
four open channel scattering in an isotropic waveguide $\omega_1=\omega_2=\omega_\perp$ using the screened Coulomb potential \cite{Saeidian2008}
are reproduced by the present approach. Note that for convergence typically 200 basis functions are used.

\section{Results and Discussion}

\vspace*{-1mm}
\subsection{Multichannel scattering in isotropic waveguides}

\begin{figure}[t]
\includegraphics[width=8.5cm] {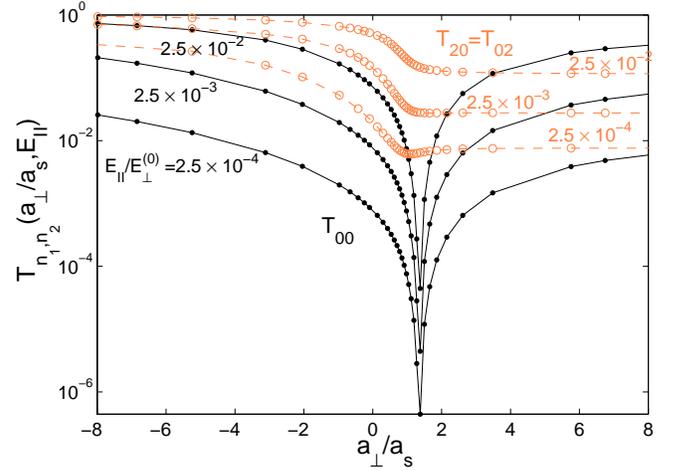}
 \caption{(color online) Dependence of the partial transmission coefficients $T_{n_1,n_2}(a_\perp/a_s,E_{\parallel})$
 on $a_\perp/a_s$ and $E_{\parallel}$ in an isotropic waveguide $\omega_1/\omega_2=1$.}\label{fig1}
\end{figure}
\begin{figure}[t]
 \includegraphics[width=8.5cm] {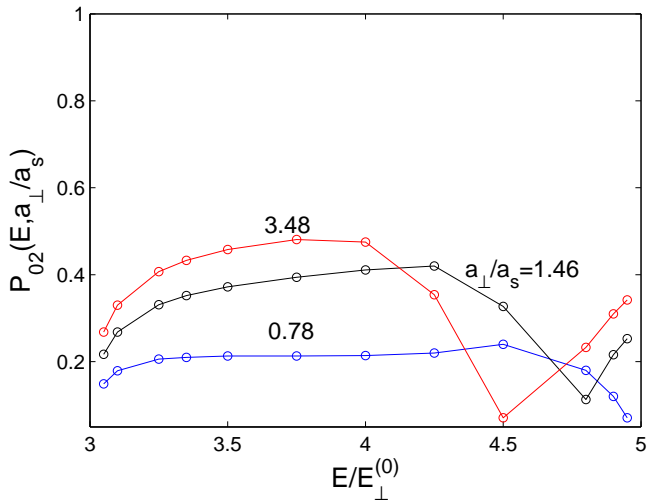}
 \caption{(color online) Transition probability $P_{02}(E,a_\perp/a_s)$ from the initial ground $n=n_1+n_2=0$ into the transversally excited states $n=n_1+n_2=2$
for a few values of $a_\perp/a_s$.} \label{fig3e}
\end{figure}

Let us first analyze atomic collisions in a transversely isotropic
waveguide $\omega_1=\omega_2=\omega_\perp$.  The
partial transmission coefficients describing the transmission in the ground $T_{00}$ and
in the first excited $T_{20}$ and $T_{02}$ scattering channels are shown in Fig.\ref{fig1} for different
longitudinal collision energies $E_{\parallel}$. Obviously the position of the CIR is stable with respect to variations of the energy
for $10^{-4}\lesssim E_{\parallel}/E_{\perp}^{(0)}\lesssim 10^{-2}$. Indeed, this position is very close to
the value $a_{\perp}/a_s=1.4603...$ predicted in the zero-energy limit with the pseudo-potential approach \cite{Olshanii1998}
and has been confirmed in subsequent numerical computations with different interatomic
potentials \cite{Bergeman2003,LowDSystems,Saeidian2008}. Hereafter we
define the position of the CIR in the ground state scattering channel as the minimum of the transmission
coefficient $T_{00}(a_\perp/a_s,E)$ which coincides with the zero of the imaginary part of the scattering amplitude
$f_{00}(a_\perp/a_s,E)$ in the region $E_\perp^{(0)}\leq E \leq E_\perp^{(2)}$
\cite{Olshanii1998,Bergeman2003,LowDSystems,Saeidian2008}.  The minimum of
the transmission coefficient $T_{20}(a_\perp/a_s)=T_{02}(a_\perp/a_s)$
in the first excited state (the position of the CIR in the excited scattering state) is at a nonzero value
and the corresponding transmission valley is therefore much less pronounced than in the case of $T_{00}(a_\perp/a_s)$
for the ground state. Moreover, the width of the CIR differs considerably for the two cases
and unlike the ground state the position of the CIR in the excited
state is strongly dependent on the energy $E_{\parallel}$. These facts, as we will
see below, are of crucial importance for the analysis of the region near the CIR. Importantly, according to (1)
the total transmission coefficient is determined in a broad neighbourhood of the resonant region $a_\perp/a_s \sim 1.4603...$
by the partial coefficients $T_{20}$ and $T_{02}$ (if $W_{20}$ and $W_{02}$ are large enough)
\begin{equation}
T(a_\perp/a_s)\approx W_{20} T_{20}(a_\perp/a_s) + W_{02}
T_{02}(a_\perp/a_s) + ... \,\,\,,
\end{equation}
which is due to the near-zero values of $T_{00}(a_\perp/a_s)$ around its minimum (see Fig.\ref{fig1}).

A note is in order here. The significant occupation of transversally excited states does not necessarily
arise due to a finite temperature in thermal equilibrium where the occupation probability is determined
by the corresponding Boltzmann distribution. Instead, the preparation of the ultracold atomic ensemble in the waveguide itself can lead to an occupation
of excited states. This could e.g. be a nonadiabatic loading process into the waveguide. The resulting
nonequilibrium state would show a redistribution of its energy between the longitudinal and transversal
degrees of freedom and might eventually thermalize, depending on the number of atoms and integrability
aspects of the underlying system \cite{Rigol}.

Atomic collisions near the CIR could possibly also lead to an increase of the excited
state population. Indeed, the increase of the atomic loss and heating near a CIR has been attributed
\cite{Haller2010} to inelastic three-body collisions, which lead to the formation of molecules while
transferring the molecular binding energy and the following energy release due to possible deexcitation of
rovibrationally excited molecules to the kinetic energy of the center of mass motion of the molecule and in particular to the
escaping third atom. Subsequent collisions of this third atom can lead to transverse excitations if its
kinetic energy exceeds the threshold value $2\hbar\omega_\perp=E_\perp^{(2)}-E_{\perp}^{(0)}$.
The binding energy of the most weakly bound molecular state in the waveguide (see Fig.2 in \cite{Bergeman2003})
exceeds already this threshold energy. Using the calculated scattering
amplitudes $f_{n_1,n_2}^{n_1',n_2'}$ we have evaluated the transition probabilities
$P_{n n'}$ \cite{Saeidian2008}
\begin{equation}
P_{n n'} =
2 \sum_{n_1',n_2'(n_1'+n_2'=2)}\frac{k_{n_1',n_2'}}{k_{n_1,n_2}}\mid
f_{n_1,n_2}^{n_1',n_2'}\mid^2\,\,\,.
\end{equation}
where $n=n_1+n_2$ and $n'=n_1'+n_2'$ .
The calculated probabilities $P_{02}(E,a_\perp/a_s)$ are presented in Fig.\ref{fig3e} for a few
values of $a_\perp/a_s$ in
the region $E_\perp^{(2)} < E < E_\perp^{(4)}$ i.e. between the first and second excited channel thresholds.
Since the probability $P_{02}(E,a_\perp/a_s)$ approaches to the values $\sim 0.2-0.4$ one might consider
collisions with the atoms accelerated due to molecules formation near CIR as a possible mechanism for the
emergence of considerable populations of the first excited states $n=2$.

Note also that hereafter we do not address the transmission coefficients $T_{n_1 n_2}$ with odd quantum
numbers $n_1=n_2=1,3,...$ because these states are not coupled with the
ground and excited states possessing even $n_1$ and $n_2$.

\subsection{Multichannel scattering in anisotropic waveguides}

\begin{figure}[t]
 \includegraphics[width=8.5cm] {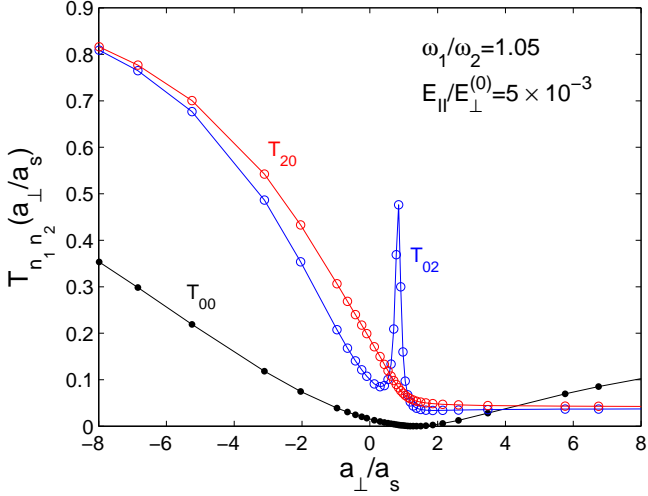}
 \caption{(color online) Partial transmission coefficients $T_{n_1,n_2}(a_\perp/a_s,E_{\parallel})$
in an anisotropic waveguide $\omega_1/\omega_2=1.05$ as functions of $a_\perp/a_s$ calculated for a near-threshold
collision energy $E_{\parallel}/E_\perp^{(0)} = 5 \times 10^{-3}$.} \label{fig2}
\end{figure}
\begin{figure}[t]
 \includegraphics[width=7.5cm] {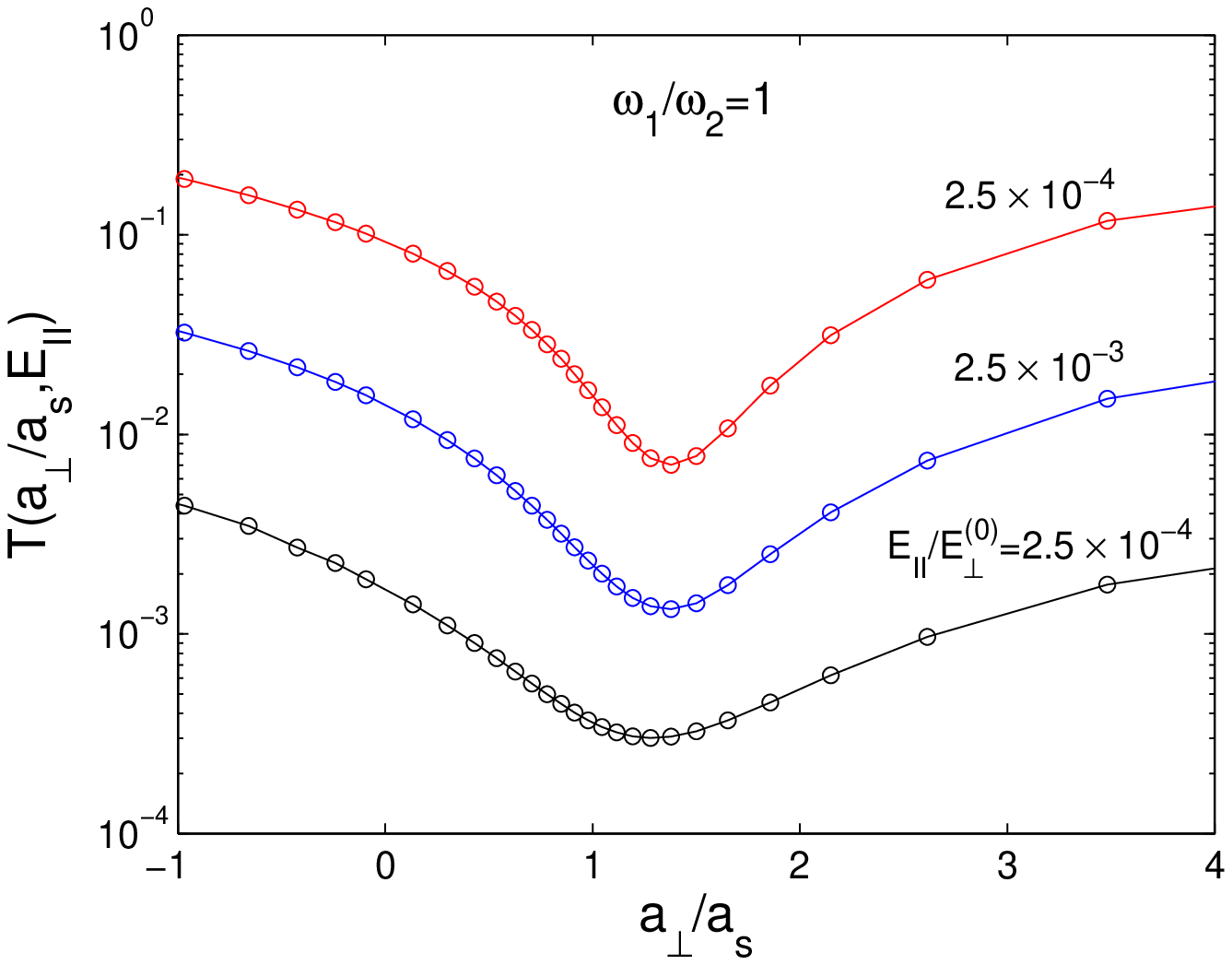}
\vspace*{3mm}

 \includegraphics[width=7.5cm] {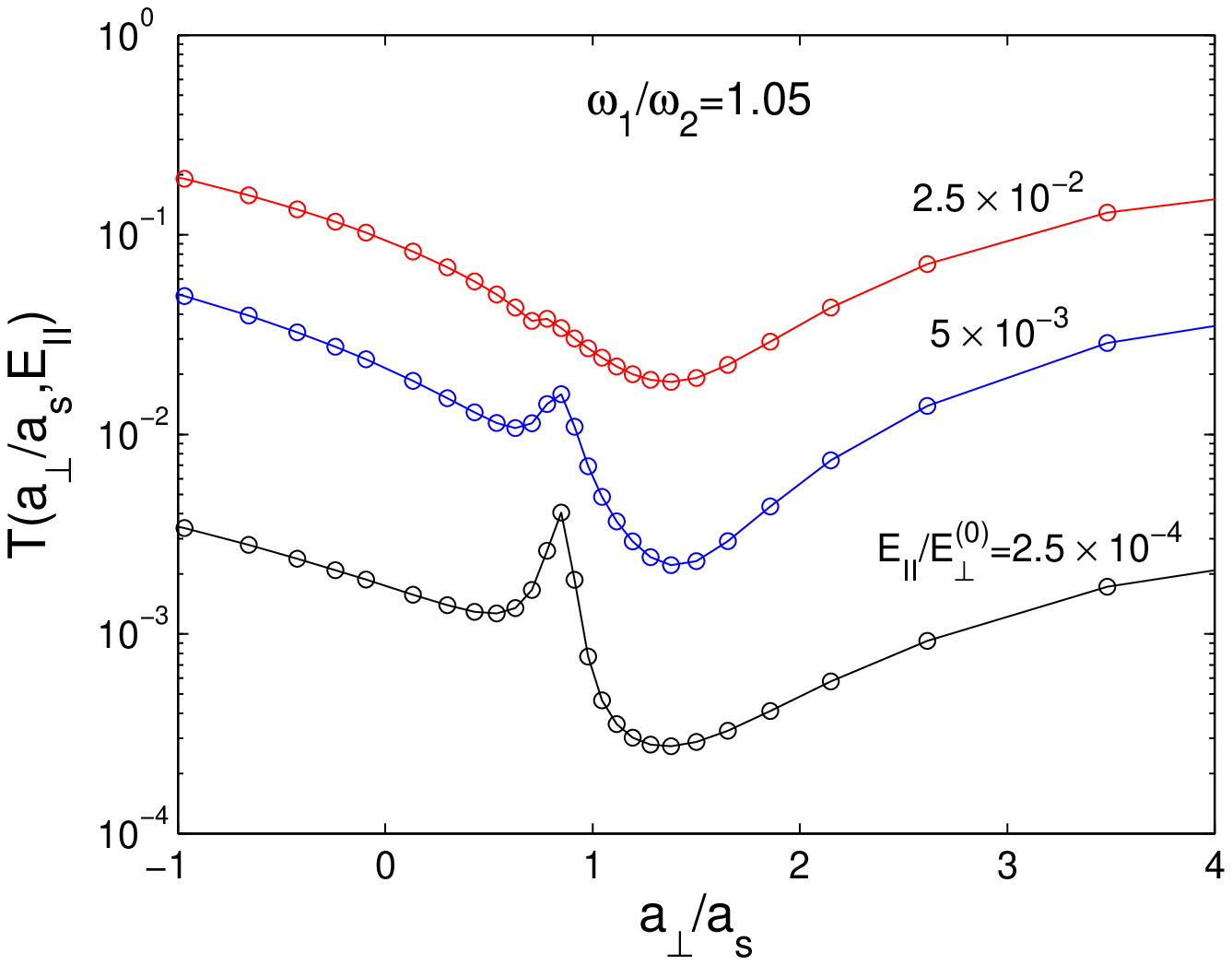}
\vspace*{3mm}

 \includegraphics[width=7.5cm] {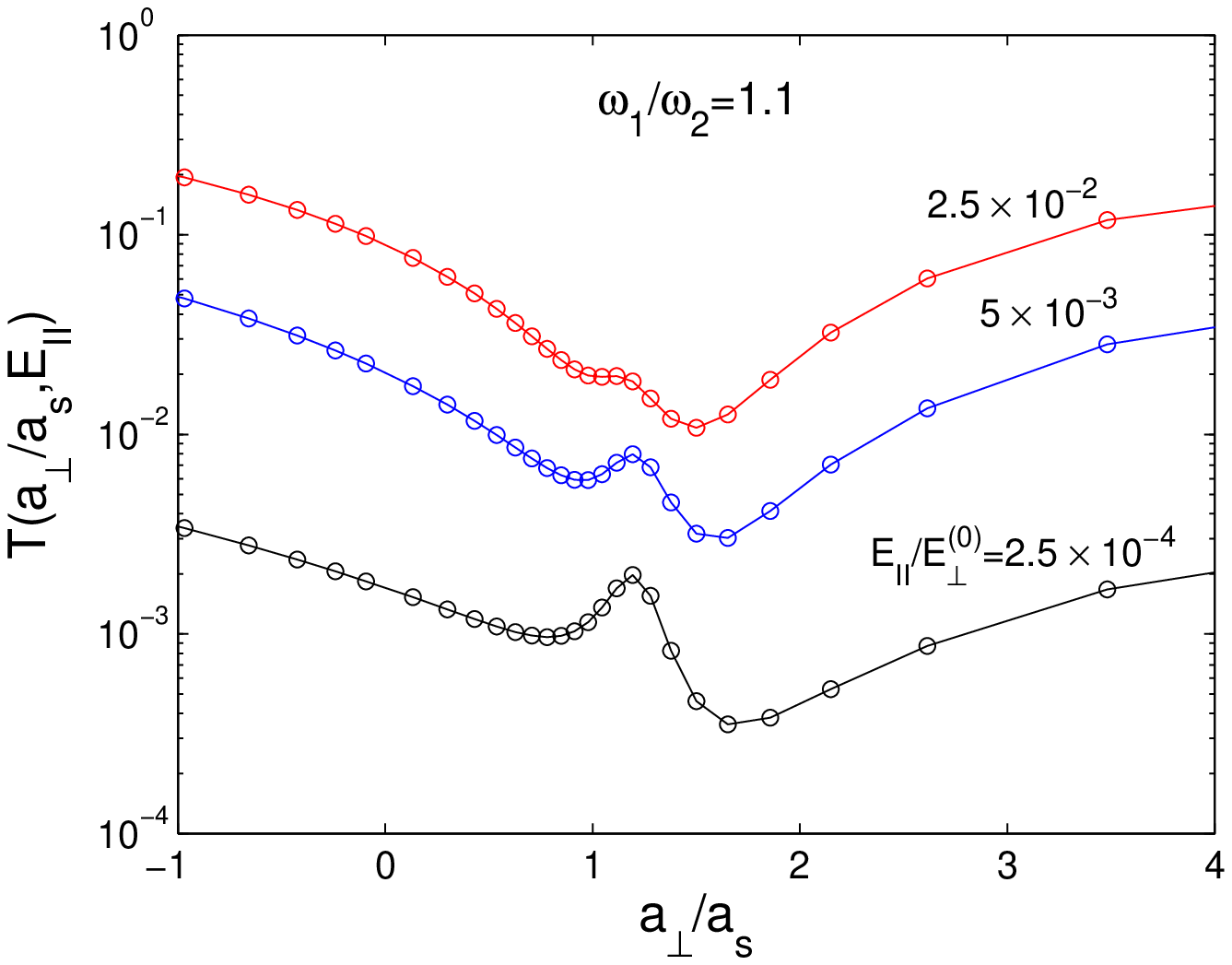}
 \caption{(color online) Total transmission coefficients $T(a_\perp/a_s,E_\parallel)$
 in the region $a_\perp /a_s \sim 1.4603...$ of the CIR for the isotropic case $\omega_1/\omega_2=1$ as well as
 for anisotropic waveguides with $\omega_1/\omega_2=1.05$ and $1.1$ calculated for $W_2/W_0=0.05$. For
 $\omega_1/\omega_2\neq 1$ the splitting of the minimum of the transmission coefficient can be observed. } \label{fig3}
\end{figure}
\begin{figure}[t]
 \includegraphics[width=8.5cm] {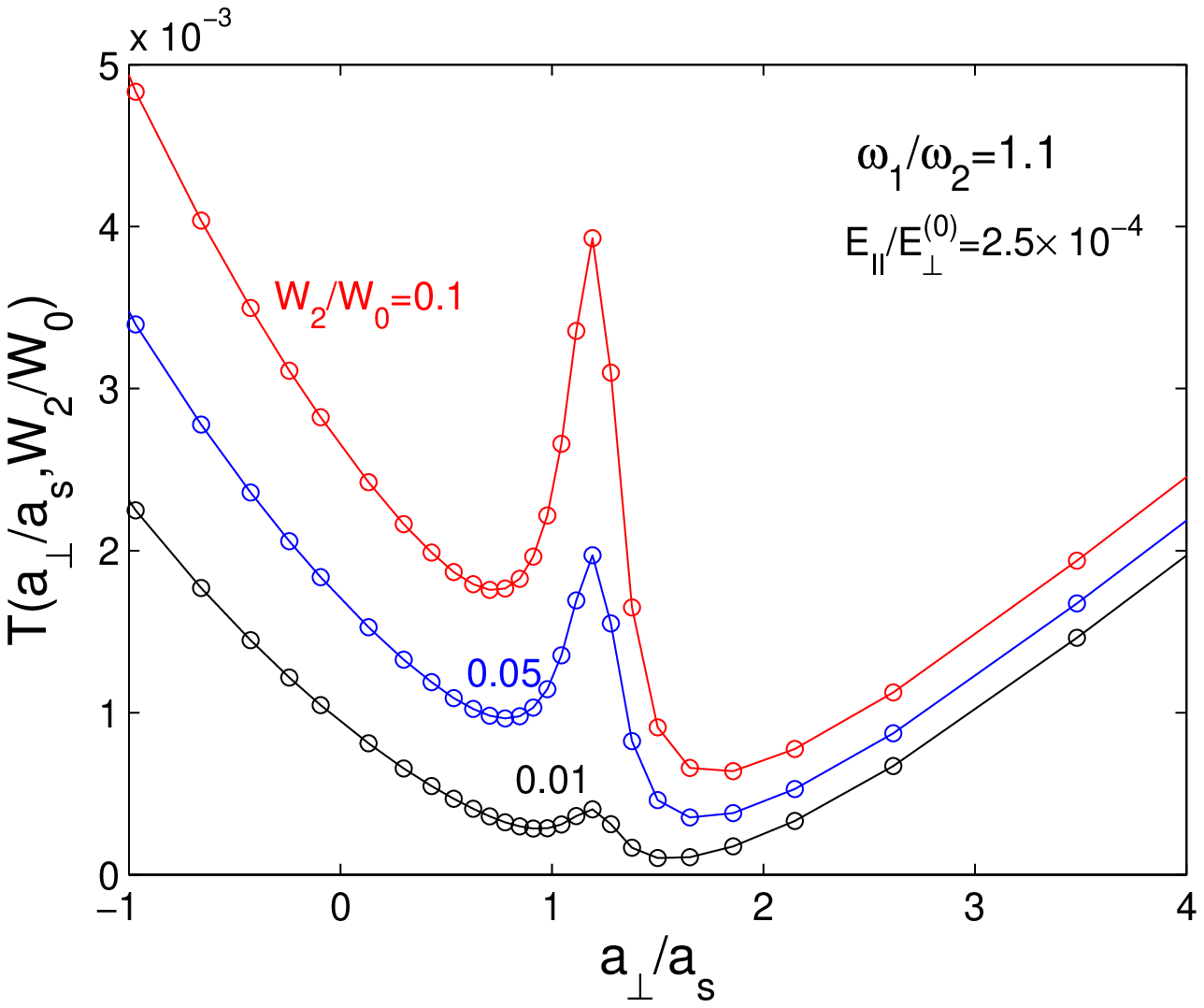}
 \caption{(color online) Total transmission coefficients $T(a_\perp/a_s,W_2/W_0)$
 in the region $a_\perp /a_s \sim 1.4603...$ of the CIR for anisotropic case $\omega_1/\omega_2=1.1$ as a function of
 relative population $W_2/W_0$ of the first excited manifold $n=2$. The splitting of the minimum of the transmission
 coefficient is still present even for only one percent population of the excited states.} \label{fig3d}
\end{figure}
We now analyze the multichannel scattering amplitude and the
corresponding transmission coefficients for anisotropic waveguides for
different values of the ratio $\omega_1/\omega_2\neq 1$. In Fig.\ref{fig2} we present the
calculated partial transmission coefficients $T_{n_1 n_2}(a_\perp/a_s)$ as a
function of $a_\perp/a_s$ for fixed values of $\omega_1/\omega_2$ and
$E_{\parallel}$. Comparing this result with the isotropic case
given in Fig.\ref{fig1} we see that the anisotropy does not change the overall behavior of the coefficients $T_{00}$ and $T_{20}$ in the region near
the CIR. However, the anisotropy splits the excited states with $n=n_1+n_2=2$ into two
components and changes the $T_{02}$ coefficient dramatically by splitting
the well of the transmission curve. As it is demonstrated in Fig.\ref{fig3}, the
effect of the splitting of the minimum of the partial transmission coefficient
$T_{02}(a_\perp/a_s)$ can also be observed in the total transmission
coefficient $T(a_\perp/a_s)$ (1) at 5\% of the relative population $W_2/W_0$ of the first exited states and
persists with varying $\omega_1/\omega_2$ and $E_{\parallel}$. Fig.\ref{fig3d} demonstrates the dependence of the
splitting on the relative population $W_2/W_0$ calculated in the region $W_2/W_0\lesssim 0.1$ compatible with the experiment
\cite{Haller2010}. The effect of the splitting of the transmission coefficient is enhanced with increasing excited state
population.

\begin{figure}[t]
 \includegraphics[width=8.5cm] {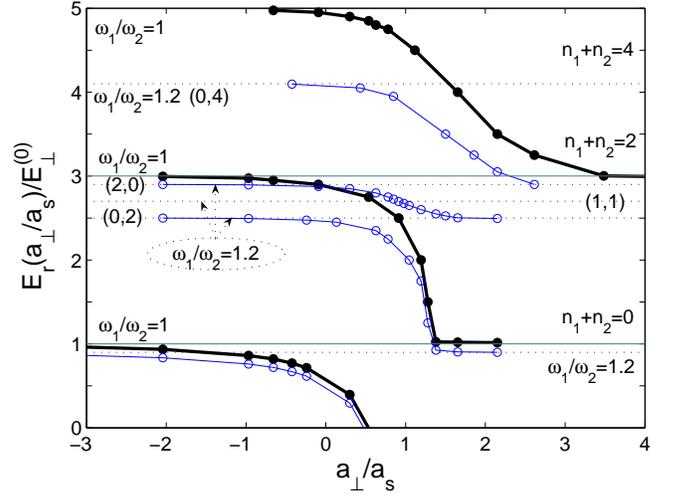}
 \caption{(color online) Illustration of the spectrum of the atomic dimer in harmonic isotropic $\omega_1/\omega_2=1$
 and anisotropic $\omega_1/\omega_2=1.2$ waveguides as a function of $a_\perp/a_s$. Below the continuum threshold
 $E_\perp^{(0)}$ the calculated binding energy of the weakly-bound state is shown. Between the thresholds $E_\perp^{(0)}$ and
 $E_\perp^{(4)}$ the calculated resonant energies $E_r$ were determined by the minimum of the transmission coefficient
 $T_{00}(a_\perp/a_s,E)$. Between the $E_\perp^{(0)}$ and $E_\perp^{(2)}$ thresholds the resonant energy $E_r$  coincides with the location
 of the zero of $Im\{f_{00}(a_\perp/a_s)\}$. Bold solid curves with the full dots correspond to the isotropic case $\omega_1/\omega_2=1$ and
 the thin curves with open circles to the anisotropic case $\omega_1/\omega_2=1.2$. Indices $(n_1,n_2)$ label the splitted sub-levels of the
 threshold energies $E_\perp^{(n_1,n_2)}(\omega_1/\omega_2=1.2)$ of the excited states with $n=n_1+n_2$. For the second excited
 threshold with $n=4$ only the lowest sub-level $(0,4)$ is shown.} \label{fig4}
\end{figure}

In the computations, the colliding energy $E_{\parallel}$ was chosen in agreement with the conditions
of the experiment \cite{Haller2010}, where $E_{\parallel}$, even at maximal heating, has remained below $k_B\times 30 nK$, i.e.
$E_{\parallel}/E_\perp^{(0)} < 30 nK/600 nK = 5\times 10^{-2}$.
It should be noted that the pronounced shift of the minimum of the $T$
coefficients with varying $E_{\parallel}$ is due to the considerable dependence of the partial coefficients $T_{02}$ and $T_{20}$
on $E_\parallel$ (see Fig.\ref{fig1}). This effect might be responsible for the shift of the maximum
of the atom loss in the experiment \cite{Haller2010} in the direction of increasing values for $a_s$. Using this assumption we can fix
$E_\parallel\backsimeq 2.5\times 10^{-4} E_\perp^{(0)}$ as being closest to the experimental conditions \cite{Haller2010} by
choosing from the calculated $T(a_\perp/a_s,E_\parallel)$ curves for $\omega_1/\omega_2=1$ the one whose position of the minimum
coincides more close with the point of maximal atomic loss in the experiment.

\vspace*{-1mm}
\subsection{Diatomic weakly-bound and resonant states in anisotropic harmonic waveguides}

To clarify the origin of the splitting of the partial coefficient $T_{02}$
we have calculated the spectrum of the near-threshold bound state and resonant
states of the atomic dimer in the confining trap as a function of
$a_\perp/a_s$ (see Fig.\ref{fig4}). The resonant energies $E_r(a_\perp/a_s)$ are
determined by the position of the minimum of the partial coefficient $T_{00}(a_\perp/a_s,E)$,
where the two-body total energy $E$ was varied between the thresholds corresponding to the ground state
$E_\perp^{(00)}=\frac{\hbar}{2}(\omega_1+\omega_2)$ and the second excited states
$E_\perp^{(n_1,n_2)}=\hbar[\omega_1(n_1+\frac{1}{2})+\omega_2(n_2+\frac{1}{2})]$ with $n=n_1+n_2=4$. These
resonant states were defined in \cite{Saeidian2008} as CIRs with non-zero
energies. The binding energy $E_B$ of the atomic dimer in the harmonic trap
with respect to the ground state threshold $E_\perp^{(0)}$ has been calculated by solving the corresponding
eigenvalue problem. 

In analyzing the results it is important to note the different dependence of the position of the CIR obtained
for the different transmission curves $T_{00}(a_{\perp}/a_s),T_{02}=T_{20}(a_{\perp}/a_s)$ in
Fig.\ref{fig1} on the collisional energy $E_{\parallel}$.
Since the resonant energy $E_r$ belonging to the CIR in the ground state strongly changes with
varying $a_\perp/a_s$  near the value $a_\perp/a_s=1.4603...$ (see Figure \ref{fig4} where a steep descent can be
observed with increasing $a_{\perp}/a_s$) the dependence of $T_{00}(a_\perp/a_s,E)$ on the energy $E_{\parallel}$
near the minimum position is very weak (see Fig.\ref{fig1}).
Opposite to this, the dependence of the resonant energy $E_r$ for the first excited states on $a_\perp/a_s$ is much smoother
(see uppermost black curve with full dots in Fig.\ref{fig4}).
This leads to a considerably stronger dependence of the minimum of $T_{20}$ and $T_{02}$ on $a_\perp/a_s$ on $E_\parallel$
(see Fig.\ref{fig1}).

In the anisotropic waveguide, the resonant curve $E_\perp^{(0)}\leq E_r(a_\perp/a_s)\leq E_\perp^{(2)}$ of the CIR of the ground
state splits into two components which are shown in Fig.\ref{fig4}. The energetically lower
(0,2)-component qualitatively repeats the behavior of the resonant energy curve (CIR) of the isotropic case.
This is why the anisotropy causes only a limited quantitative change for the $T_{00}$ coefficient. However, the behavior of the
energetically upper (2,0)-component is different (see Fig.\ref{fig4}). The resonance curve $E_r$ of this component is flat for the complete parameter region
of $a_\perp/a_s$ which leads to a strong change with respect to the energy dependence of the
$T_{02}$ coefficient, including in particular a strong change of the corresponding positions of the minima (maxima) with respect to
$a_{\perp}/a_s$ . The behavior of the $T_{20}$ coefficient is determined by
the energetically lowest resonant curve $E_r$ emerging from the $n=4$ threshold in the isotropic case which is deformed only
slightly by the anisotropy. This is why the difference between $T_{20}(\omega_1/\omega_2\neq 1)$ and $T_{20}(\omega_1/\omega_2= 1)$ is
significantly smaller when compared to the case of the $T_{02}$ coefficient.

The above model could potentially also explain the appearance of additional CIRs with further increase of the
anisotropy $\omega_1/\omega_2\neq 1$ as seen in \cite{Haller2010}. Increasing $\omega_1/\omega_2\neq 1$
the energetical distance between the sub-levels characterized by $n=2, 4, ...$ decreases. This leads to a considerable increase
of the populations $W_{n_1,n_2}$ of higher excited states and in particular to an increase of the contribution of
these states to the total transmission coefficient $T$ (see eq.(1)).

\section{Mechanism of the CIR splitting in anisotropic harmonic waveguide}

The key for the understanding of the mechanism of the splitting of the CIR under the action of an anisotropic harmonic trap
is the diatomic spectrum of the weakly-bound and resonant states in the harmonic waveguide given in Fig.6. 
It also explains why the previous considerations in the zero-energy limit near the ground state threshold 
did not provide any splitting of the CIR defined as the singularity of $g_{1D}= lim_{E_{\parallel}\rightarrow 0}(k_{00}Ref_0/Imf_0)$ 
\cite{Peng2010,Zhang2}. 

Actually, so far it was implicitly supposed that the resonant energy curve $E_r$ behaves linearly as it crosses the ground state threshold $n=0$ for
$a_{\perp}/a_s=1.4603$ and becomes subsequently a weakly-bound state for $a_{\perp}/a_s > 1.4603$ (see, for example, Fig.2 in \cite{Bergeman2003}). 
Therefore, it was natural to expect that in an anisotropic waveguide $\omega_1-\omega_2=\Delta\omega\rightarrow 0$ leading to the splitting of the first excited threshold 
$E_{\perp}^{(2,0)}- E_{\perp}^{(0,2)}=\hbar\Delta\omega$ (see Fig.6 in the present paper and Fig.1(b) in \cite{Haller2010}) 
the resonant curve $E_r$ will also split into two components $E_r^{(2,0)}$ and $E_r^{(0,2)}$ crossing the ground state threshold 
at the points $a_{\perp}/a_s^{(2,0)}$ and $a_{\perp}/a_s^{(0,2)}$ with the separation $a_{\perp}/a_s^{(0,2)}-a_{\perp}/a_s^{(2,0)}$
proportional to the threshold splitting $\hbar\Delta\omega$. These expectations were confirmed in the experiment by measuring the distance between the 
maxima of the atomic loss in the anisotropic waveguide which was interpreted as $a_{\perp}/a_s^{(0,2)}-a_{\perp}/a_s^{(2,0)}$ (see Fig.3(c) in \cite{Haller2010}). 

However, our extension of the calculation of the resonant energy $E_r$ to the region  $a_{\perp}/a_s > 1.4603$ has shown a
strongly nonlinear behavior of the curve $E_r$ (see Fig.6) while shifting the point where the resonant curve $E_r$ crosses the ground state threshold 
to the value $a_{\perp}/a_s \rightarrow +\infty$. This means that at the point $a_s\rightarrow +0$ we observe a complete rearrangement of the
spectrum. The first resonant state $E_r$ becomes a new weakly-bound state once we cross this point and equivalently for the higher excited 
resonant states which convert into each other.

This behavior, i.e. the rearrangement of the spectrum of the two-body system at the point  $a_s\rightarrow +0$, remains in the anisotropic waveguide (see Fig.6). 
This is why the splitting of the singularity of the function $g_{1D}= lim_{E_{\parallel}\rightarrow 0}(k_{00}Ref_0/Imf_0)$ was not observed near
the point $a_{\perp}/a_s=1.4603$ in the anisotropic harmonic waveguide \cite{Peng2010,Zhang2} but one can observe the splitting of the minimum
in the effective transmission coefficient $T$ (eq.(1)) due to the quasi-crossing of the resonant curves $E_r^{(0,2)}$ and 
$E_r^{(2,0)}$ leading to the splitting of the minimum in the $T_{02}$ coefficient (see Fig.3)
qualitatively equal to the width of the quasi-crossing near $a_{\perp}/a_s=1.4603$. 
This width is proportional to the first excited threshold splitting $\hbar\Delta\omega$ in the presence of the anisotropy
and in good agreement with the experimental value for the splitting of the maxima of the atom loss \cite{Haller2010}. 

\vspace*{-2mm}
\section{Conclusions}

Our investigation and following analysis demonstrates
that multichannel scattering in anisotropic harmonic waveguides can lead to a splitting of the confinement-induced resonance.
A necessary ingredient is a population of at least a few percent of
the transversally excited states which can certainly occur via e.g. a nonadiabatic loading process of the atoms into the waveguide. 

There are several ways to improve our suggested model which would help further clarifying the behaviour of the splitting effect
with varying parameters and depending on the initial preparation of the atomic ensemble.
First, one would have to take into account the velocity distribution (distribution over the
collision energy $E_\parallel$) in the longitudinal direction \cite{Griesmaier2009}.
Shifts and splittings of the CIR due to anharmonicities of the trap \cite{Kestner2010,Saenz2011} and the influence of the
inelastic channel of molecule formation have also to be determined. 

It is known that the anharmonicity of the trap couples the relative and center-of-mass motion of the
colliding atoms and can lead to additional anharmonicity-induced resonances (AIRs) due 
to the removal of the degeneracy of the center-of-mass and relative motions \cite{Kestner2010,Saenz2011}.
However, the AIRs are supposed to be much narrower than the CIRs and, 
as a consequence, more difficult to be detected experimentally because of the relatively weak anharmonic coupling 
with respect to the interatomic interaction \cite{Peano2005,Kestner2010}. 
As a conclusion, the observation of these AIRs presumably represents a challenging experimental problem.
Explicitly suppressing the anharmonicity of the waveguide
could lead to a discrimination between the two mechanisms causing a splitting of the CIRs.

\vspace*{-2mm}
\section{Acknowledgements}

We thank Dr. E.Haller for fruitful discussions. The authors acknowledge financial support by the Deutsche Forschungsgemeinschaft and the Heisenberg-Landau Program.
V.S.M. thanks the Zentrum f\"ur Optische Quantentechnologien of the University of Hamburg for the warm hospitality.

\bibliographystyle{apsrev}

\begin{references}
\bibitem{Olshanii1998} M. Olshanii, Phys. Rev. Lett. {\bf 81}, 938 (1998).
\bibitem{Kinoshita2004} T. Kinoshita, T. Wenger, and D. S. Weiss, Science \textbf{305}, 1125 (2004).
\bibitem{Paredes2004} B. Paredes, A. Widera, V. Murg, O. Mandel, S. F\"olling, I. Cirac, G.V. Shlyapnikov, T.W. H\"ansch, and E. Bloch, Nature \textbf{429}, 277 (2004).
\bibitem{Haller2009} E. Haller, M. Gustavsson, M.J. Mark, J.G. Danzl, R. Hart, G. Pupillo, and H.C. N\"agerl, Science \textbf{325}, 1224 (2009).
\bibitem{Gunter05} K. G\"unter, T. St\"oferle, H. Moritz, M. K\"ohl, and T. Esslinger, Phys. Rev. Lett. {\bf 95}, 230401 (2005).
\bibitem{Saeidian2008} S. Saeidian, V.S. Melezhik, and P. Schmelcher, Phys. Rev. A {\bf 77}, 042721 (2008).
\bibitem{Melezhik2009} V.S. Melezhik and P. Schmelcher, New J. Phys. {\bf 11}, 073031 (2009)
\bibitem{Peano2005} V. Peano, M. Thorwart, C. Mora, and R. Egger, New J. Phys. {\bf 7}, 192 (2005)
\bibitem{Melezhik2006} J.I. Kim, V.S. Melezhik and P. Schmelcher, Phys. Rev. Lett. {\bf 97}, 193203 (2006)
\bibitem{Haller2010} E. Haller, M.J. Mark, R. Hart, J.G. Danzl, L. Reichs\"ollner, V. Melezhik, P. Schmelcher, and H.C. N\"agerl, Phys. Rev. Lett. {\bf 104}, 153203 (2010).
\bibitem{Peng2010} S.G. Peng, S.S. Bohloul, X.J. Liu, H. Hu, and P.D. Drummond, Phys. Rev. A {\bf 82}, 063633 (2010).
\bibitem{Zhang2} W. Zhang and P. Zhang, Phys.Rev. A {\textbf{83}}, 053615 (2011)
\bibitem{Kestner2010} J.P. Kestner and L.M. Duan, New J. Phys. {\bf 12}, 053016 (2010)
\bibitem{Saenz2011} S. Sala, P.-I. Schneider, and A. Saenz, arXiv:1104.1561.
\bibitem{Bergeman2003} T. Bergeman, M. G. Moore, and M. Olshanii, Phys. Rev. Lett. {\bf 91}, 163201 (2003).
\bibitem{LowDSystems} E. Tiesinga, C.J. Williams, H.H. Mies, and P.S. Julienne, Phys. Rev. A {\bf 61}, 063416 (2000); V.A. Yurovsky, Phys. Rev. A {\bf 71}, 012709 (2005);
V.S. Melezhik, J.I. Kim, and P. Schmelcher, Phys. Rev. A {\bf 76}, 053611 (2007).
\bibitem{Melezhik1991} V.S. Melezhik, J. Comp. Phys. {\bf 92}, 67 (1991)
\bibitem{Melezhik2003} V.S. Melezhik
 and C.-Y. Hu, Phys. Rev. Lett. {\bf 90}, 083202 (2003)
\bibitem{Rigol} M. Rigol, Phys. Rev. Lett. {\textbf{103}}, 100403 (2009)
\bibitem{Griesmaier2009} A. Griesmaier, A. Aghajani-Talesh, M. Falkenau, J. Sebastian, A. Greiner, and T. Pfau, J. Phys. B \textbf{42}, 145306 (2009).
\end{references}

\end{document}